\documentstyle[epsf,preprint,aps]{revtex}
\def\beq{\begin{eqnarray}}
\def\eed{\end{eqnarray}}
\begin{document}
\draft

\title{Schwinger-Boson Mean-Field Theory of Mixed-Spin
Antiferromagnet $L_2BaNiO_5$}

\author{Yun Song and Shiping Feng}

\address{Department of Physics, Beijing Normal University,
Beijing, 100875, China}

\date{\today }
\maketitle

\begin{abstract}
The Schwinger-boson mean-field theory is used to study the
three-dimensional antiferromagnetic ordering and excitations in compounds
$L_2BaNiO_5$, a large family of quasi-one-dimensional mixed-spin antiferromagnet.  
To investigate magnetic properties of these compounds, we introduce a 
three-dimensional mixed-spin antiferromagnetic Heisenberg model 
based on experimental results for the crystal structure of $L_2BaNiO_5$.
This model can explain the experimental discovery of 
coexistence of Haldane gap and  
antiferromagnetic long-range order below N\'{e}el temperature.
Properties such as the low-lying excitations, magnetizations of $Ni$
and rare-earth ions, N\'{e}el temperatures of different 
compounds, and  the behavior of Haldane gap below the N\'{e}el temperature 
are investigated within this model, and the results 
are in good agreement with neutron scattering experiments.
\end{abstract}
\vskip5mm


\section{Introduction}

The existence of Haldane gap \cite{Haldane} in the magnetic excitation
spectrum has kept the integer-spin one-dimensional (1D) Heisenberg
antiferromagnet to be one of the most interesting subjects in condensed
matter physics during the past twenty years. The integer-spin Heisenberg
antiferromagnetic (AF) chain should have a singlet ground state,
exponentially decaying correlations, and a quantum gap, which have been
confirmed by numerous theoretical 
\cite{Affleck,Nightingale,Nomura,Sakai,Meshkov} and experimental 
\cite{Morra,Ma,Zheludev,Buttrey} studies. 
Kennedy and Tasaki have proved that the appearance of Haldane gap in the spin-1 AF chain corresponds to the breaking of a hidden $Z_2\times Z_2$ symmetry \cite{Kennedy}.
The recent discovery of coexistence
of Haldane gap and AF long-range order (AF LRO) in rare-earth 
compounds $L_2BaNiO_5$ with $L$=$Y,Nd,Sm,Eu,Gd,Tb,Dy,Ho,Er,$ and $Tm$
\cite{Buttrey,Matres,Sachan,Zheludev2,Yokoo,Raymond} has offered 
an opportunity
to investigate the effect of staggered field on the Haldane chain. The
polarized and unpolarized inelastic neutron scattering 
experiments on $L_2BaNiO_5$ \cite
{Zheludev2,Raymond} present that, on the one hand, there are AF interactions
between spins of rare-earth and $Ni$ ions, which lead to the
three-dimensional (3D) AF LRO with N\'{e}el temperature ($T_{N}$) ranging from $20-70K%
$ \cite{Matres,Sachan}. On the other hand, these compounds are also
characterized by the presence of $NiO_{6}$
octahedron chains along $a$-axis.
As is anticipated, above the N\'{e}el temperature, 
energy gap has been found in the excitation propagating along 
$Ni$-chain. 
Also the Haldane gap is discovered to
persist in the 3D AF phase and
 increase with decreasing temperature below $T_N$ 
\cite{Zheludev2,Yokoo,Raymond}.

Recently several theoretical works have focused on the quasi-1D
Haldane system $L_2BaNiO_5$ \cite{Maslov,Lou,Takushima}. To describe the
coexistence of Haldane gap and  
AF LRO, the nonlinear sigma model in a static staggered magnetic
field has been introduced by Maslov and Zheludev 
\cite{Maslov}. This system has
been studied numerically within the density matrix renormalization
group method by adopting the model of an 1D spin-$1$ Heisenberg chain
in a static field \cite{Lou}. In the above studies, the reason for
introducing the static staggered field is based on the assumption that there
are 3D directly AF exchange interactions between the spins of rare-earth
ions, which lead to the AF LRO below $T_{N}$. However, 
it has been pointed out by some experiments \cite{Matres,Sachan,Zheludev2} 
that spins of $Ni$ ions also play an 
important role in forming AF LRO, and the
two-dimensional mixed-spin Heisenberg model with $s_1=1$ and $s_2=1/2$
chains staked alternatively has also been adopted for this system \cite
{Takushima}. 

Depending on the experimental finding of crystal structure,
we introduce a more suitable 3D mixed-spin 
AF Heisenberg model to describe the magnetic
properties of compounds $L_{2}BaNiO_{5}$. 
In this model we define $s_{1}=1$ for the spins of 
$Ni$ ions, and investigate different members of this family by changing
the spin value $s_{2}$ of rare-earth ions correspondingly.
We use the Schwinger-boson mean-field (SBMF) \cite{Arovas,Auerbach,Sarker} theory to study this 3D mixed-spin model. 
The SBMF theory  has been 
successful to the study of spin-integer Heisenberg chains
\cite{Arovas,Auerbach}, and has also been extended to cases with magnetic
ordering by identifying the magnetization
with the Bose condensation of the Schwinger bosons \cite{Arovas,Auerbach,Sarker}.
Within this model,
we could explain the experimental discovery of 
coexistence of Haldane gap and AF LRO.
The interactions between spins of $Ni$
and rare-earth ions are proved to be very important in forming the AF LRO.
Properties such as the magnetization, Haldane gap and N\' {e}el temperatures
are also discussed 
for cases with different $s_{2}$. 
Comparing our results with neutron scattering experiments, 
we find that our findings could explain some experimental discoveries.

The paper is organized as follows: In Sec. II we introduce 
the 3D mixed-spin Heisenberg model and Schwinger-boson mean-field theory. 
Our results of the magnetic properties of $L_2BaNiO_5$ compounds are 
presented in Sec. III. Finally, we conclude our findings in Sec. IV.

\section{Mixed-Spin Heisenberg Model and Schwinger-Boson Mean-field Theory}

The crystal structures of $L_2BaNiO_5$ compounds have been investigated
extensively by neutron scattering 
experiments \cite{Matres,Sachan,Zheludev2}. 
They belong to the orthorhombic system, having approximate cell
Parameters around $a=3.8\AA$, $b=5.8\AA$ and $c=11.3\AA$.
 In all members of this compounds, as figure 1(b) shown, strings of distorted $NiO_6$
octahedron share apical oxygen and form Haldane chains along the $a$-axis.
The intrachain $Ni$-$O$-$Ni$ AF superexchange
coupling is about $200$-$300K$ \cite{Matres,Sachan,Zheludev2},
which is the strongest magnetic interaction in the compound. 
While in the $bc$ plane, as figure 1(a) shown, 
there is one oxygen between the $Ni$ and rare-earth ($L$)ions,
and thus 
the $Ni$-$O$-$L$ interactions establish links between individual $Ni$ chains.
This coupling is of the order of tens of Kelvin ($10$-$30K$). Figure 1(a) also
shows that there are $L$-$O$-$L$ AF interactions between rare-earth ions, which is of the order of $1K$ or less.

Based on the experimental discovery of the crystal structure of compounds 
$L_2BaNiO_5$, we introduce a 3D mixed-spin AF Heisenberg model expressed
by the Hamiltonian 
\begin{eqnarray}
H =J_1\sum_{i,\eta _{a}}S_{1,i}\cdot S_{1,i+\eta _{a}}+J_2\sum_{i,\eta _{b},\eta
_{c}}S_{1,i}\cdot S_{2,i+\eta _{b}+\eta _{c}}  
+J_3\sum_jS_{2,j}\cdot S_{2,j+\widehat{c_{0}}}~,
\end{eqnarray}
where $S_{1}$ and $S_{2}$ are the spin operators of $Ni$ and rare-earth ions
respectively. $\eta _{b}$ and $\eta _{c}$ denote a sum over nearest-neighbor (NN)
bonds in the $bc$ plane, and so is $\eta _{a}$ for the NN bonds along the 
$a$-axis. The AF superexchange couplings along the $Ni$ chain, between $L$ and $%
Ni$ ions, and among the rare-earth $L$ ions are represented as $J_1$, $J_2$,
and $J_3$ respectively as figure 1(a) shown.

The Schwinger-boson theory introduces $S_i=\frac 12b_{i\alpha }^{+}\sigma
_{\alpha ,\beta }b_{i\beta }$ ($\alpha $ (or $\beta $)$=\uparrow ,\downarrow 
$)\cite{Arovas,Auerbach}, and the spin degrees of freedom are mapped to the
boson degrees of freedom. Meanwhile the original spin Hilbert space corresponds
to a boson Hilbert subspace in which $b_{i\uparrow }^{+}b_{i\uparrow
}+b_{i\downarrow }^{+}b_{i\downarrow }=2s_i$. These constraints to the boson
Hilbert space are imposed in the Hamiltonian (1) by introducing two
kinds of Lagrangian multipliers $\lambda _{1}(i)$ and $\lambda _{2}(j)$. In addition,
we define the bond operators as $Q_a(i,\eta _a)=b_{1i\uparrow
}b_{1(i+\eta _a)\downarrow }-b_{1i\downarrow }b_{1(i+\eta _a)\uparrow }$,
$Q_h(i,\eta _b,\eta _c)=b_{1i\uparrow }b_{2(i+\eta _b+\eta_c)\downarrow }
-b_{1i\downarrow }b_{2(i+\eta _b+\eta _c)\uparrow }$, and 
$Q_c(j,\widehat{c_{0}})=b_{2j\uparrow }b_{2(j+\widehat{c_{0}})\downarrow }
-b_{2j\downarrow }b_{2(j+\widehat{c_{0}})\uparrow }$. 
The Hamiltonian (1) can be rewritten as

\begin{eqnarray}
H &=&-\frac{1}{2}J_1\sum_{i,\eta _a}\left\{ Q_a^{+}(i,\eta _a)Q_a(i,\eta
_a)-2s_1^2\right\}-\frac 12J_3\sum_j\left\{ Q_c^{+}(j,\widehat{c_{0}})Q_c(j,\widehat{c_{0}}%
)-2s_2^2\right\}    \nonumber \\
&&-\frac 12J_2\sum_{i,\eta _b,\eta_c}\left\{ Q_h^{+}(i,\eta _b,\eta _c)Q_h(i,\eta
_b,\eta _c)-2s_1s_2\right\} \nonumber \\
&&+\sum_i\lambda _1(i)\left \{ b_{1i\uparrow }^{+}b_{1i\uparrow
}+b_{1i\downarrow }^{+}b_{1i\downarrow }-2s_1\right \}  
+\sum_j\lambda _2(j)\left \{ b_{2j\uparrow }^{+}b_{2j\uparrow
}+b_{2j\downarrow }^{+}b_{2j\downarrow }-2s_2\right \}.
\end{eqnarray}

Next, we make a Hartree-Fock decomposition of Eq. (2) by taking the average
values of the bond operators and Lagrange multipliers to be uniform and
static as $\langle Q_a(i,\eta _a)\rangle =Q_a$, $\langle Q_h(i,\eta _b,\eta
_c)\rangle =Q_h$, $\langle Q_c(j,\widehat{c_{0}})\rangle =Q_c$, $\langle \lambda
_1(i)\rangle =\lambda _1$, and $\langle \lambda _2(j)\rangle =\lambda _2$.
Under the Fourier transformation, we obtain the following mean-field
Hamiltonian in the momentum space

\begin{eqnarray}
H^{MF} &=&\sum_k2Z_k\left\{ b_{1k\uparrow }b_{1(-k)\downarrow
}-b_{1k\downarrow }b_{1(-k)\uparrow }+h.c\right\}  
+\sum_k4D_k\left\{ b_{1k\uparrow }b_{2(-k)\downarrow }
-b_{1k\downarrow}b_{2(-k)\uparrow }+h.c\right\}  \nonumber \\
&&+\sum_k2\chi _k\left\{ b_{2k\uparrow }b_{2(-k)\downarrow
}-b_{2k\downarrow }b_{2(-k)\uparrow }+h.c\right\}  
+\sum_{k\sigma }\left\{ 4\lambda _1b_{1k\sigma }^{+}b_{1k\sigma
}+8\lambda _2b_{2k\sigma }^{+}b_{2k\sigma }\right\}  \nonumber \\
&&+2N\left \{J_1Q_a^2+4J_2Q_h^2+J_3Q_c^2-4\lambda _1s_1
+2J_1s_1^2+8J_2s_1s_2+2J_3s_2^2-8\lambda _2s_2\right \},
\end{eqnarray}
where $Z_k=J_1Q_a\cos (k_za_{0})$, $\chi _k=J_3Q_c\cos (k_xc_{0})$, 
and $D_k=2J_2Q_h\cos (k_xc_{0})\cos (k_yb_{0})$.

Diagonalizing $H^{MF}$ in Eq. (3) by the
Bogoliubov transformation, we obtain

\begin{eqnarray}
H^{MF} =2\sum_k\left\{ E_k^{+}\left( \alpha _{1k\sigma }^{+}\alpha _{1k\sigma
}+\alpha _{2k\sigma }\alpha _{2k\sigma }^{+}\right)  
+E_k^{-}\left( \beta _{1k\sigma }^{+}\beta _{1k\sigma }+\beta _{2k\sigma
}\beta _{2k\sigma }^{+}\right) \right \}+E_0
\end{eqnarray}
with

\begin{eqnarray}
E_k^{\pm } &=&\sqrt{\frac{ A\pm \sqrt{A^2-4B} }2}  \nonumber \\
A &=&\lambda _1^2+4\lambda _2^2-\left( Z_k^2+\chi _k^2+2D_k^2\right)  
\nonumber \\
B &=&4\lambda _1\lambda _2\left( \lambda _1\lambda _2-D_k^2\right) -\chi
_k^2\lambda _1^2-4Z_k^2\lambda _2^2+\left( \chi _kZ_k-D_k^2\right) ^2 
\nonumber \\
E_0&=&2N\left \{J_1Q_a^2+4J_2Q_h^2+J_3Q_c^2 
+2J_1s_1^2+8J_2s_1s_2+2J_3s_2^2\right\} \nonumber \\
&&-8N\lambda _1\left( s_1+\frac 12%
\right) -16N\lambda _2\left( s_2+\frac 12%
\right).
\end{eqnarray}

Furthermore, the mean-field free energy is given by

\begin{eqnarray}
F^{MF} =\frac 8\beta \sum_k \ln \left \{\left( 2\sinh \left( \frac{\beta
E_k^{+}}2\right) \right) 
+\ln \left( 2\sinh \left( \frac{\beta E_k^{-}}2%
\right) \right)\right\} +E_0.  
\end{eqnarray}

The mean-field equations are obtained by differentiating $F^{MF}$ with
respect to the parameters $Q_a$, $Q_h$,  $Q_c$, $\lambda _1$, and 
$\lambda _2$ as

\begin{eqnarray}
\frac{\partial F^{MF}}{\partial Q_a}=\frac{\partial F^{MF}}{\partial Q_h}=%
\frac{\partial F^{MF}}{\partial Q_c}=
\frac{\partial F^{MF}}{\partial \lambda
_1}=\frac{\partial F^{MF}}{\partial \lambda _2}=0.
\end{eqnarray}
Thus we obtain five self-consistent equations to determine the average values
of the bond operators and Lagrange multipliers.

\section{Mean-Field solutions}
In this section, we present the solutions of the SBMF theory.
The experimental discovery of coexistence of 
Haldane gap and AF LRO
below $T_{N}$ are obtained by our study. 
As equation (5) shown, there are two branches of 
the magnetic excitations $E_{k}^{+}$ and $E_{k}^{-}$,
which have quite different behaviors
below $T_{N}$. 
On the one hand, the magnetic excitation 
$E_{k}^{+}$ has an energy gap in the whole 
temperature region, and below $T_{N}$ this energy gap increases 
with decreasing temperature. 
On the other hand, within the temperature
region from zero to $T_{N}$,
$E_{k}^{-}$ keeps to be gapless and
has its minimal value $E^{-}_{k}=0$ at $k=0$. 
Under this condition, the Schwinger-boson condensation occurs and 
leads to the AF LRO in this system. In our calculation, we introduce the 
Schwinger-boson condensation into the self-consistent mean-field
equations, and obtain the temperature dependence of the 
magnetizations of $Ni$ ($M_{1}$) and rare-earth ($M_{2}$) ions below $T_{N}$
respectively. 
The magnetizations $M_{1}$ and $M_{2}$ are expressed as      
\begin{eqnarray}
M_{1}&=&\langle S_{1}^{Z}\rangle =\frac{1}{2N}\sum_{i}\langle 
b_{1i\uparrow }^{+}b_{1i\uparrow }-b_{1i\downarrow }^{+}b_{1i\downarrow
}\rangle   \nonumber \\
M_{2}&=&\langle S_{2}^{Z}\rangle =\frac{1}{2N}\sum_{j}\langle 
b_{2j\uparrow }^{+}b_{2j\uparrow }-b_{2j\downarrow }^{+}b_{2j\downarrow
}\rangle. 
\end{eqnarray}
We choose the AF superexchange 
interactions as $J_{1}=J$, $J_{2}=0.1J$ and $J_{3}=0.01J$ ($J>0$) according 
to the experimental studies of the magnetic properties of 
compounds $L_{2}BaNiO_{5}$ \cite{Matres,Sachan,Zheludev2}. 
The AF interactions between $Ni$ ions has been estimated as
$2J=200-300K$ by the neutron scattering experiments 
\cite{Matres,Zheludev2}.
To make it simple and clear, in our calculation we choose $J=100K$.

The neutron scattering experimental 
results of $Nd_{2}BaNiO_{5}$
obtained by Yokoo {\it {et al.}} \cite{Yokoo} 
are shown in Fig. 2(a) and 2(b) (open circles).  
The solid lines in Fig. 2(a) and 2(b) 
show fits to the above experimental results by 
our numerical calculations for the case with $s_{2}=3/2$ respectively.   
The temperature dependence of magnetizations
$M_{1}/s_{1}$ and $M_{2}/s_{2}$
for the cases of $s_{2}=1/2,1,3/2,2,5/2$, and $3$
are shown in figure 2(c) and 2(d) respectively. 
As temperature increases, the thermal fluctuation in the system 
becomes stronger, and
the magnetizations decrease rapidly
and drop to zero at the N\' {e}el temperatures. 
Therefore, we could also determine the N\' {e}el temperatures $T_{N}$ for all 
members of the $L_{2}BaNiO_{5}$ family through our calculation.

In figure 3, we plot the N\'{e}el temperature 
$T_{N}$ as a function of spin value $s_{2}$ (filled circles 
and dotted line).
For comparison, the N\'{e}el temperatures of compounds
$Ho_{2}BaNiO_{5}$ ($T_{N}=53K$) \cite{Matres},  
$Nd_{2}BaNiO_{5}$ ($T_{N}=48K$) \cite{Sachan}
and $Pr_{2}BaNiO_{5}$ ($T_{N}=24K$)\cite{Zheludev2} obtained by 
the neutron scattering experiments
are also shown in Fig. 3 (filled triangles).
Our theoretical results roughly agree with the experimental results.
We also obtain that the N\' {e}el temperature $T_{N}$ increases monotonously
with the increasing of $s_{2}$.
As the spin value of the rare-earth ions $s_{2}$ equal to 
$1/2$ and $3$, we
get that the minimize and maximum N\' {e}el temperature are 
$T^{Min}_{N}\approx 0.207J$ and $T^{Max}_{N}\approx 0.872J$
respectively. 
Our result of the N\' {e}el temperature
region is approximately from $20.7K$ to $87.2K$ when $J=100K$,
which is in good agreement with the experimental
estimation of $T_{N}$ region $20-70K$ \cite{Matres,Sachan}.    

The two branches of the magnetic excitation $E^{+}_{k}$ and $E^{-}_{k}$ 
represent the spin fluctuations along the $a$-axis and within
$bc$ plane respectively. In the exactly one-dimensional case ($J_{2}=J_{3}=0$), 
the excitation  $E^{-}_{k}$ vanishes and the energy gap of
$E^{+}_{k}$ is just the Haldane gap of AF Heisenberg chain, which is closely related to the breaking of a hidden $Z_2\times Z_2$ symmetry \cite{Kennedy}.
We obtain that the coexistence of Haldane gap
and AF LRO below $T_{N}$ is a common feather of 
all members of compounds $L_{2}BaNiO_{5}$ except
$Y_{2}BaNiO_{5}$. The neutron scattering experiments
have discovered that, below the N\' {e}el temperature, the Haldane gap
increases as the temperature decreases
\cite{Zheludev2,Yokoo,Raymond}.
The temperature dependences of the Haldane gap $\Delta$ for the
cases of $s_{2}=1/2,1,3/2,2,5/2$ and $3$ are also investigated by 
the SBMF theory and the behaviors are shown in 
figure 4(a) respectively. 
The temperature dependence of the energy gap $\Delta$
obtained by our calculation is found to agrees with the experimental
discovery.
 
Our calculation also implies that the effect of the staggered magnetization on the magnetic excitation is to widen the Haldane gap.
Below $T_{N}$, the effective internal magnetic field 
$H_{eff}$ imposed on the Haldane chain is assumed approximately as
the magnetization $M_{1}$ of $Ni$ ions.
This field is found to increase with the decreasing of temperature 
and the increasing of spin value $s_{2}$ 
because of the thermal fluctuation being weakened and 
the AF ordering being enhanced.
In figure 4(b), we plot the energy gap $\Delta$ of zero temperature 
as a function of $H_{eff}$, and 
we find that $H_{eff}$ has strong effect to widen the Haldane gap.

In addition, based on the 3D mixed-spin model, we obtain that AF LRO below 
the N\' {e}el temperature is not constructed only by the rare-earth ions. 
Our results support the suggestion that $Ni$ ions also play an important role in forming the AF LRO.  
We plot in figure 5 the N\'{e}el temperature as a function of AF coupling $J_2$ in the cases with $s_2=1/2$, $J_1=J$, and $J_3=0.01J$. 
In compounds $L_{2}BaNiO_{5}$, the effective interactions between individual $Ni$ chains rely on the AF coupling $J_2$ between $Ni^{2+}$ and rare-earth $L^{3+}$.  
We obtain that the N\'{e}el temperature rises rapidly with the increase of $J_2$ as shown in figure 5, so the coupling $J_2$ are important in forming the AF LRO. 
Besides, as $J_{2}=0$, there are no interactions between $Ni$ chains and thus the N\'{e}el temperature drops to zero. 
The Haldane excitation energy $E^{+}_k$ in 3D and 1D cases are shown in figure 6(a) and 6(b) respectively. 
Here we choose $s_2=1/2$, $J_1=J$, $J_2=0.01J$, $J_3=0.01J$, and $k_x=k_y=0$ for 
the 3D case, and obtain that the corresponding N\'{e}el temperature is $T_{N}=0.206J$.
To compare the behaviors of magnetic excitations below and above the N\'{e}el temperature, we study two conditions of $T=0.1$ (solid lines) and $T=0.4J$ (dotted lines) respectively for both 1D and 3D cases. 
In the 3D case, we find, in figure 6(a), that the energy gap below the N\'{e}el temperature is obviously bigger than that above the N\'{e}el temperature, which is in opposition with the behavior for pure 1D case (shown in figure 6(b)).
In addition, we obtain that the thermal fluctuation has strong effect to destroy the 3D spin correlations in the compounds, as a result the behaviors in 3D case is the same as that in 1D case when temperature is above the N\'{e}el temperature. 

\section{Summary}

In conclusion, we have introduced a 3D mixed-spin AF Heisenberg model  
based on the experimental results of the crystal 
magnetic structure of compounds $L_2BaNiO_5$, and studied this model with
the SBMF theory.
The experimental discovery of coexistence of Haldane gap
and AF LRO below $T_{N}$ has been deduced by our calculation. 
Properties such as the low-lying excitations, magnetizations of $Ni$
and rare-earth ions, N\'{e}el temperatures of different 
members of this family and behavior of Haldane gap below $T_{N}$ have also been 
investigated within this model. 

We have obtained two branches of the magnetic excitations $E_{k}^{+}$ 
and $E_{k}^{-}$, of which $E_{k}^{+}$ has an energy gap and
$E_{k}^{-}$ is gapless below $T_{N}$.
The theoretical result of the N\' {e}el temperature
region is approximately from $20.7K$ to $87.2K$,
which is in good agreement with the experimental
estimation of the region $20-70K$.
We have also obtained that Haldane gap increases with decreasing temperature,
and the effect of the magnetization is to widen the Haldane gap.
Our results are in good agreement 
with the experimental discoveries.
Our findings also support the suggestion that the AF LRO below 
$T_{N}$ is not constructed only by the rare-earth ions,
and $Ni$ ions also play an important role in forming
the AF LRO.

\acknowledgments We thank Dr. Xintian Wu for helpful discussions. This work
was supported by the National Natural Science Foundation of China under Grant
Nos. 10125415, 10074007, and 90103024.

\begin{figure}[ht]
\caption{(a) Structural relation between $Ni$ and rare-earth sites in the $bc$ plane,
and (b) $Ni$ and apical oxygen form the Haldane chain along the $a$-axis. 
} 
\end{figure}
\vglue 0.5cm

\begin{figure}[ht]
\caption{ The temperature dependence of the ordered moment on the $Ni$ (a) and 
$Nd$ (b) sites in $Nd_{2}BaNiO_{5}$ sample taken from Ref. 14 (open circles), and the solid lines show fits to the experimental discovery by 
our numerical results of the case with $s=3/2$ respectively;   
Magnetizations $M_{1}/s_{1}$ (c) and $M_{2}/s_{2}$ (d) as functions of temperature below $T_{N}$ for cases of 
$s_{2}=1/2,1,3/2,2,5/2$, and $3$ respectively.
} 
\end{figure}
\vglue 0.5cm 

\begin{figure}[ht]
\caption{ N\'{e}el temperatures of the compounds
$L_{2}BaNiO_{5}$ for the cases with different spin value $s_{2}$.
The circles represent the theoretical results, and the triangles are 
experimental findings of compounds $Nd_{2}BaNiO_{5}$, $Pr_{2}BaNiO_{5}$ and
$Ho_{2}BaNiO_{5}$ taken from Ref. 11-13. 
} 
\end{figure}
\vglue 0.5cm 

\begin{figure}[ht]
\caption{ (a) The temperature dependence
of Haldane gap for cases of $s_{2}=1/2,1,3/2,2,5/2$ and $3$ 
below the N\'{e}el temperatures respectively;
(b) The Haldane gap as a function of
the effective internal magnetic field $H_{eff}$.
} 
\end{figure}
\vglue 0.5cm 

\begin{figure}[ht]
\caption{N\'{e}el temperature as a function of AF coupling $J_{2}$
between $Ni^{2+}$ and $L^{3+}$ ions of case $s_2=1/2$, $J_1=J$, and
$J_3=0.01J$.
}
\end{figure}
\vglue 0.5cm
 
\begin{figure}[ht]
\caption{The branch of Haldane excitation $E^{+}_k$ as a function 
of $k_z$ in (a) 3D with $s_2=1/2$, $J_1=J$, $J_3=0.01J$ 
and (b) pure 1D cases. The solid lines show results of
temperature $T=0.1J$, and the dotted lines present results
of $T=0.4J$.
} 
\end{figure}
\vglue 0.5cm
 

\vskip -0.5cm
\begin{thebibliography}{99}
\bibitem{Haldane}  F.~D.~M.~Haldane, Phys. Lett. {\bf 93A}, 464 (1983);
                Phys. Rev. Lett. {\bf 50 }, 1153 (1983).

\bibitem{Affleck}  I.~Affleck, J. Phys. Condens. Matter {\bf 1}, 3047 (1989).

\bibitem{Nightingale}  M.~P.~Nightingale and H.~W.~J.~Bl\H {o}te, Phys. Rev. B 
              {\bf 33}, 619 (1986).

\bibitem{Nomura}  K.~Nomura, Phys. Rev. B {\bf 40}, 2421 (1989).

\bibitem{Sakai}  T.~Sakai and M.~Takahashi, Phys. Rev. B {\bf 42}, 1090
              (1990).

\bibitem{Meshkov}  S.~V.~Meshkov, Phys. Rev. B {\bf 48}, 6167 (1993).

\bibitem{Morra}  R.~M.~Morra, W.~J.~L.~Buyers, R.~L.~Armstrong, and 
               K.~Hirakawa, Phys. Rev. B {\bf 38}, 543 (1988).

\bibitem{Ma}  S.~Ma, C.~Broholn, D.~H.~Reich, B.~J.~Sternlieb,
              and R.~W.~Erwin, Phys. Rev. Lett. {\bf 69}, 3571 (1992).

\bibitem{Zheludev}  A.~Zheludev, S.~E.~Nagler, S.~M.~Shaprio, L.~K.~Chou,
               D.~R.~Talham, and M.~W.~Meisel, Phys. Rev. B 
              {\bf 53}, 15004 (1996).

\bibitem{Buttrey}  D.~J.~Buttrey, J.~D.~Sullivan, and A.~L.~Rheingold, 
               J. Solid State Chem. {\bf 88}, 291 (1990).

\bibitem{Kennedy}  T.~Kennedy and H.~Tasaki, Phys. Rev. B {\bf 45}, 304 (1992);
                Rev. Math. Phys. {\bf 6}, 887 (1994). 

\bibitem{Matres}  E.~Garc\' {i}a-Matres, J.~Rodr\' {i}guez-Carvajal,
               J.~L.~Mart\' {i}nez, A.~Salinas-S\' {a}chez, 
               and R.~S\' {a}ez-Puche, Solid. State. Commun. 
              {\bf 85}, 553 (1993).

\bibitem{Sachan}  V.~Sachan, D.~J.~Buttrey, J.~M.~Tranquada and G.~Shirane,
               Phys. Rev. B {\bf 49}, 9658 (1994).

\bibitem{Zheludev2}  A.~Zheludev, J.~M. Tranquada, and T.~Vogt and 
               D.~J.~Buttrey, Phys. Rev. B {\bf 54}, 6437 (1996).

\bibitem{Yokoo}  T.~Yokoo, {\it {et al.}}, Phys. Rev. B {\bf 58}, 14424 (1998).

\bibitem{Raymond}  S.~Raymond, T.~Yokoo, A.~Zheludev, S.~E.~Nagler, 
               A.~Wildes, and J.~Akimitsu, Phys. Rev. Lett. 82, 2382 (1999).

\bibitem{Maslov}  S.~Maslov and A.~Zheludev, Phys. Rev. B {\bf 57}, 68
              (1998); Phys. Rev. Lett. {\bf 80}, 5786 (1998)

\bibitem{Lou}  J.~Lou, X.~Dai, S.~Qin, Z.~Su, and L.~Yu, Phys. Rev. B 
              {\bf 60}, 52 (1999).

\bibitem{Takushima}  Y.~Takushima, A.~Koga, and N.~Kawakami, Phys. Rev. B
              {\bf 61}, 15189 (2000).

\bibitem{Arovas}  D.~P.~Arovas and A.~Auerbach, Phys. Rev. B {\bf 38}, 
               316 (1988); D.~Yoshioka, J. Phys. Soc. Jpn. {\bf 58}, 
               32 (1989).

\bibitem{Auerbach}  A.~Auerbach, {\it Interacting Electrons and Quantum 
               magnetism}, (Springer-Verlag, 1994).

\bibitem{Sarker}  S.~Sarker, C.~Jayaprakash, H.~R.~Krishnamurthy 
               and M.~Ma, Phys. Rev. B {\bf 40}, 5028 (1989);
               C.~L.~Kane, P.~A.~Lee, T.~K.~Ng, B.~Chakraborty,
               and N.~Read, Phys. Rev. B {\bf 41}, 2653 (1990).

\end{thebibliography}
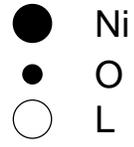
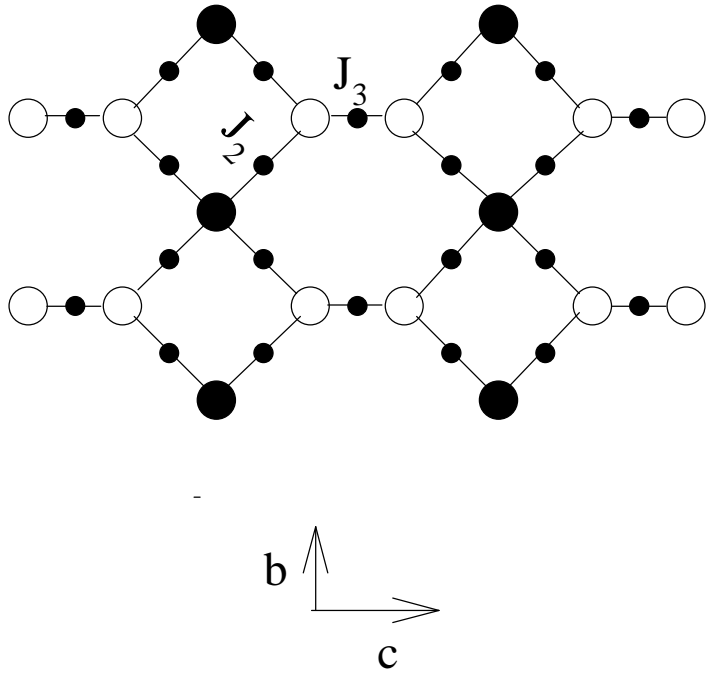
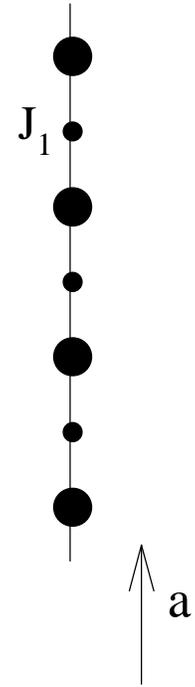
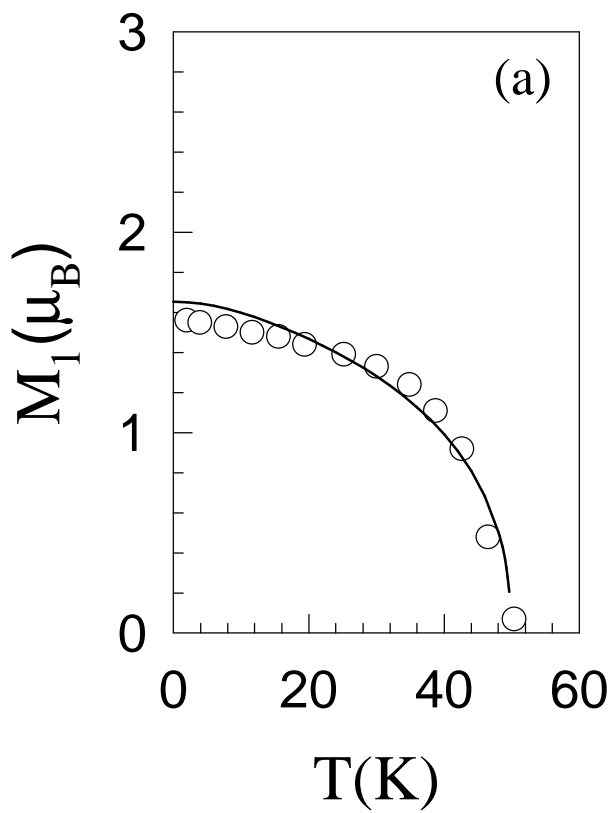
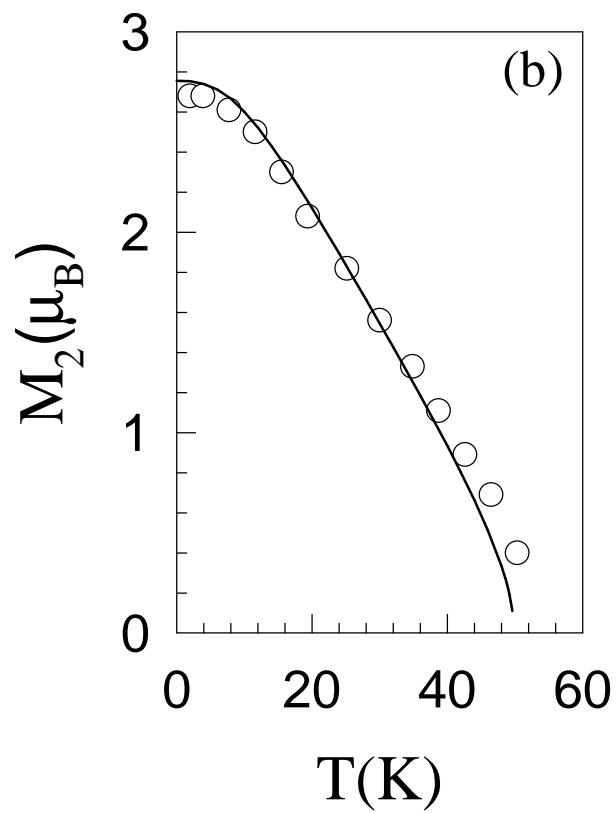
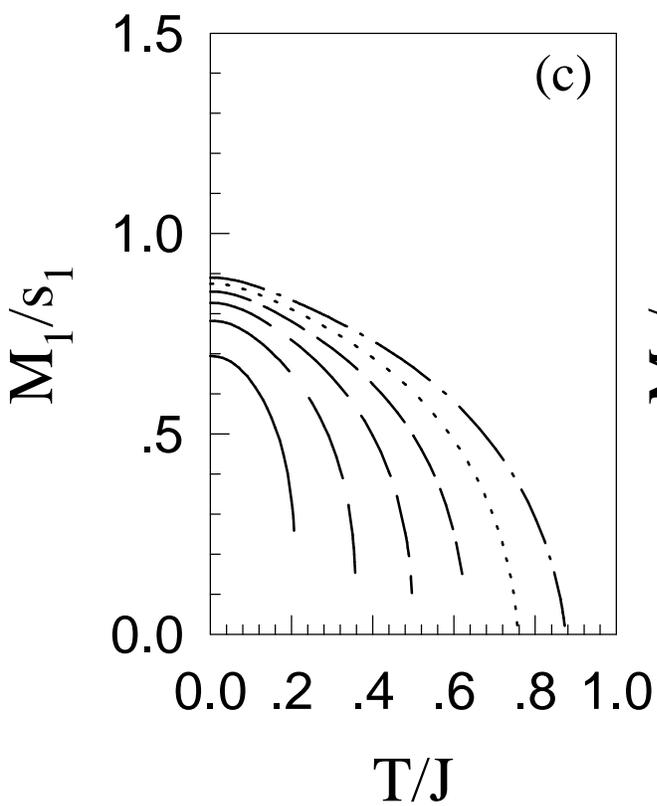
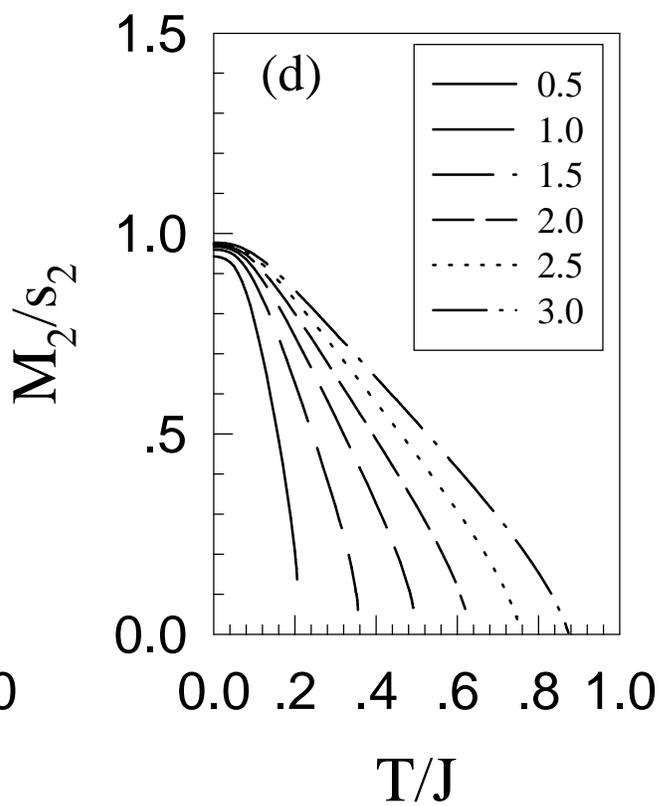
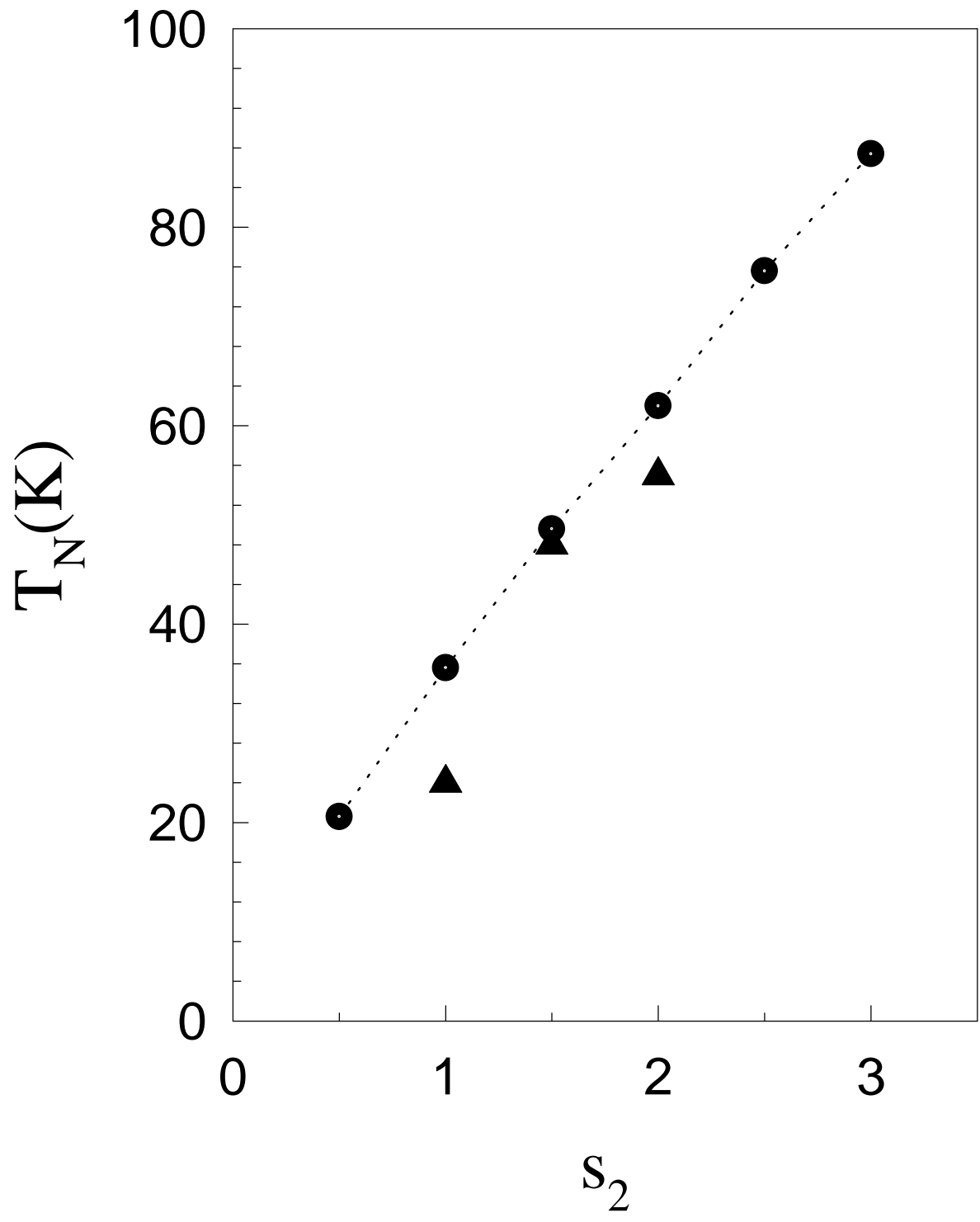
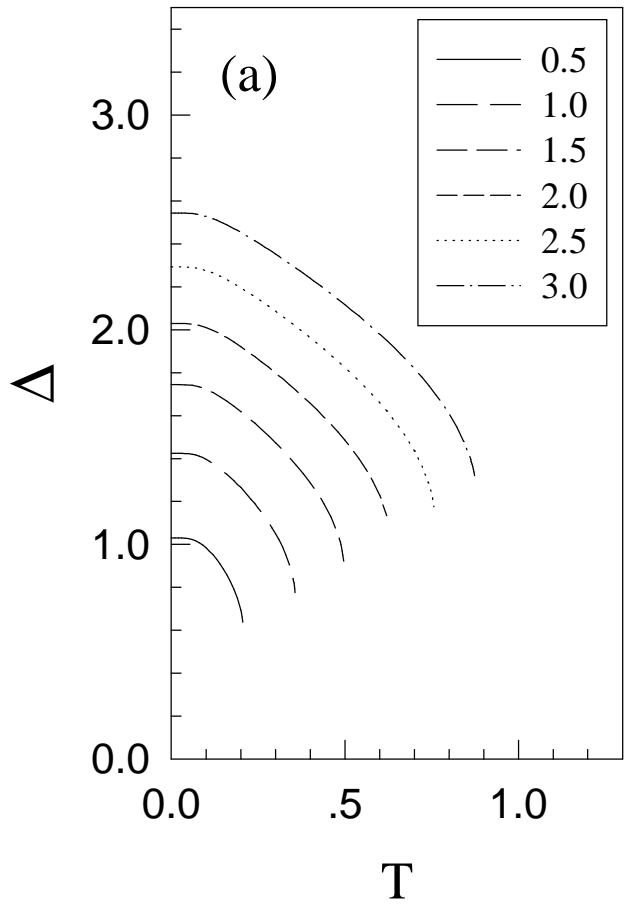
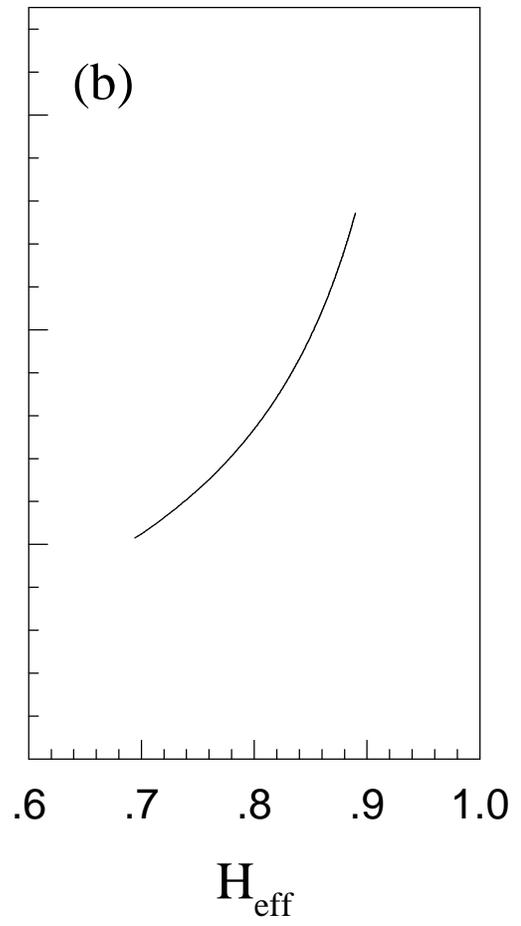
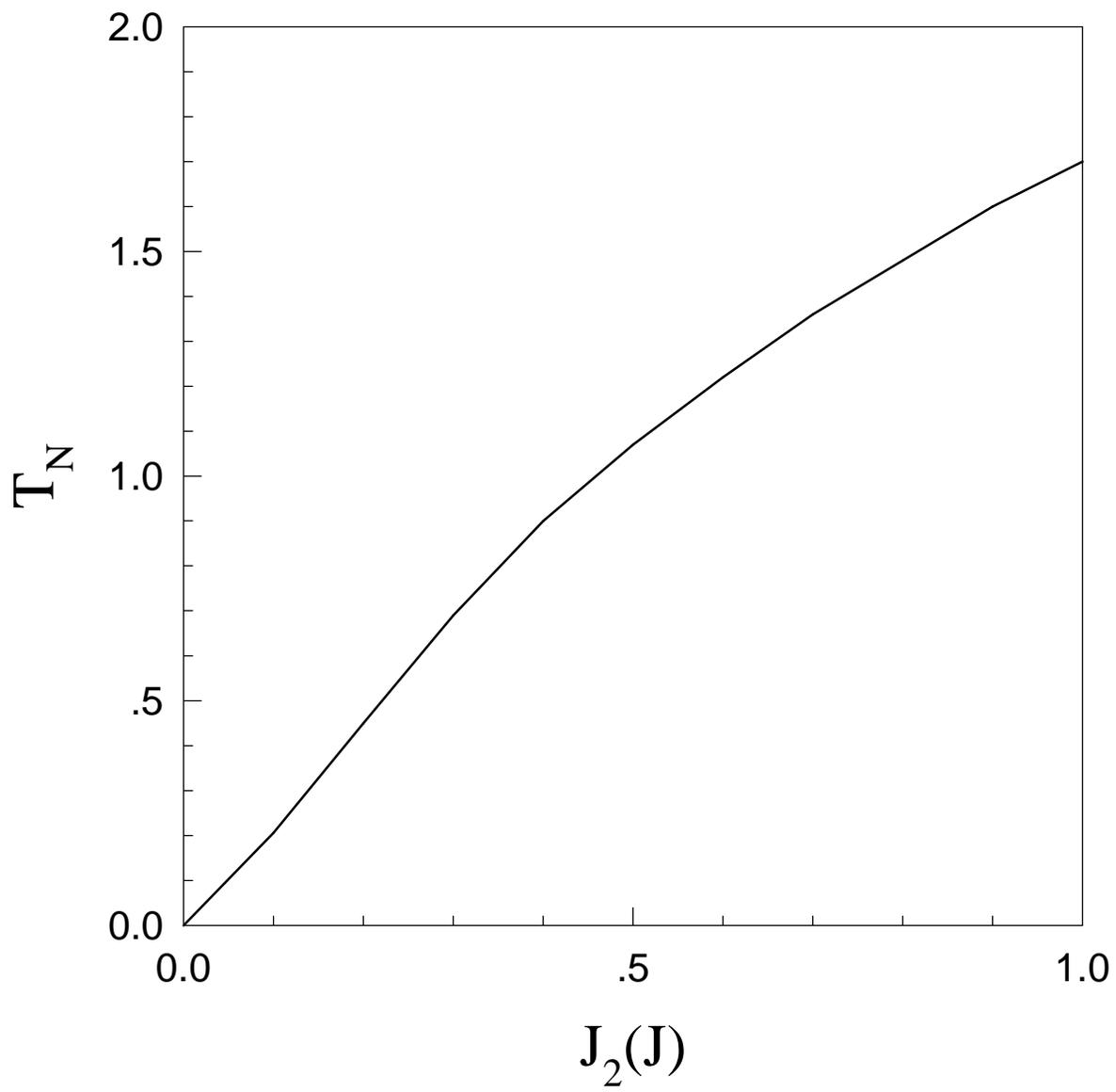
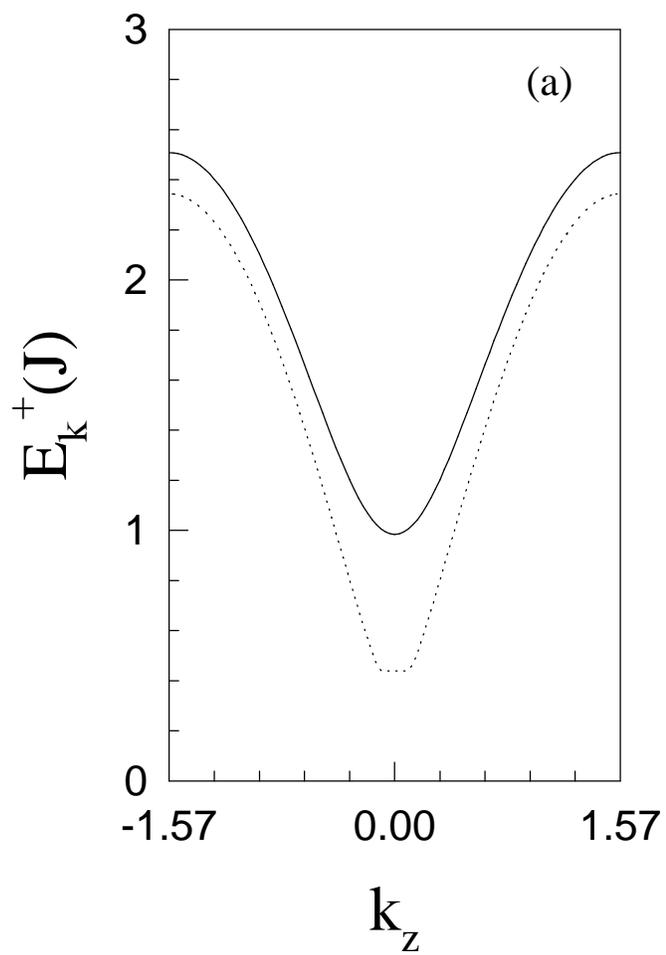
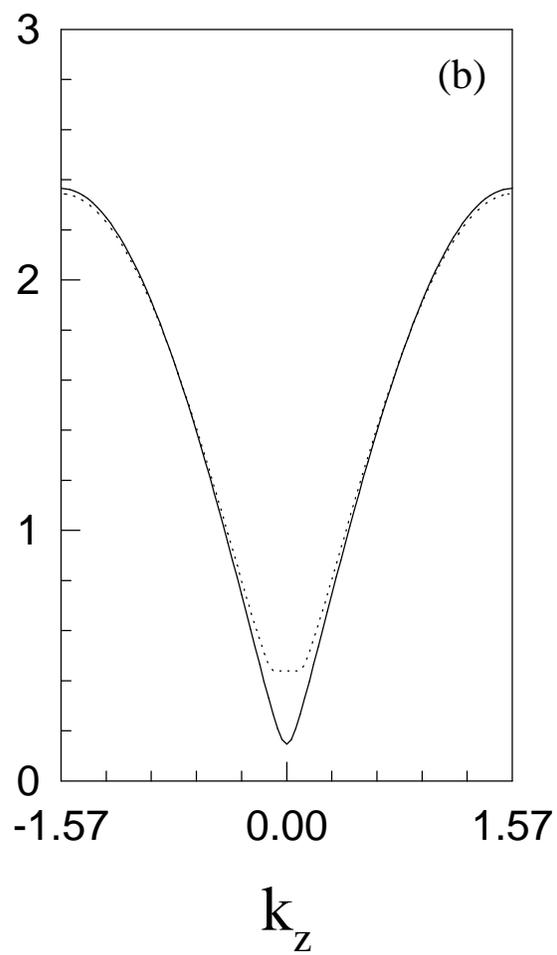
\end{document}